\documentclass[aps,prl,twocolumn,showpacs,floatfix,amsmath,amssymb]{revtex4}
\usepackage{graphicx}
\newcommand{\beq}{\begin{equation}}
\newcommand{\eeq}{\end{equation}}
\newcommand{\bea}{\begin{eqnarray}}
\newcommand{\eea}{\end{eqnarray}}
\begin{document}
\title{Fluctuation of the Correlation Dimension and the Inverse 
Participation Number at the Anderson Transition}
\author{Imre Varga}
\affiliation{Fachbereich Physik, Philipps--Universit\"at Marburg,
Renthof 6, D-35032 Marburg, Germany}
\affiliation{Elm\'eleti Fizika Tansz\'ek,
Budapesti M\H uszaki \'es Gazdas\'agtudom\'anyi Egyetem,
H-1521 Budapest, Hungary}
\date{\today}
\begin{abstract}
The distribution of the correlation dimension in a power law band
random matrix model having critical, i.e. multifractal, eigenstates
is numerically investigated. It is shown that their probability
distribution function has a fixed point as the system size is varied 
exactly at a value obtained from the scaling properties of the typical 
value of the inverse participation number. Therefore the state-to-state
fluctuation of the correlation dimension is tightly linked to the 
scaling properties of the joint probability distribution of the eigenstates. 
\end{abstract}
\maketitle{}

It is already well established \cite{MJ,rev,MEM} that the eigenstates at
the localization--delocalization transition driven by disorder are
multifractals characterized by a set of generalized dimensions, $D_q$.
This is reflected both in the statistical properties of each
eigenstate and in the joint probability distribution of the
eigenvector components. In the first case the
multifractal analysis is performed on each eigenstate using the usual
box--counting algorithm, while in the second case the scaling
properties of the distribution function are analyzed.

The generalized dimensions are associated to the scaling of the $q$th 
moment of the eigenstates, $\psi_{\alpha}(\mathbf{r})$. These moments are 
also termed as generalized inverse participation numbers 
\cite{MJ,rev}
\beq
\label{iqdq}
I_q=
\left\langle\int\,d^dr\,|\psi_{\alpha}(\mathbf{r})|^{2q}\right\rangle_{\alpha}
\sim L^{-D_q(q-1)},
\eeq
where $L$ is the linear dimension of the system. The inverse of $I_2$, 
for example, roughly equals the number of nonzero wave function 
components, therefore it is a good and widely accepted measure of the 
extension of the states. For delocalized states $I_2\sim 1/N$, while 
for localized ones $I_2\sim O(1)$. At criticality it scales with $L$ 
with exponent $D_2$ (\ref{iqdq}). This scaling may be found for each 
state $\psi_{\alpha}(\mathbf{r})$ but then the corresponding exponents 
fluctuate. 

Another property of multifractality is related to the probability
overlap of eigenstates with energy separation substantially exceeding 
the mean level separation. In particular $D_2$ describes these
density correlations: hence the name of correlation dimension. It has
been shown that for multifractal eigenstates these correlations decay
slowly \cite{FM} no matter how sparse these states are. The role of
$D_2$ is even more important as it has been related to the level
compressibility at criticality \cite{CLK,VK} that is a remarkable
relation that has been corroborated at least for the case of 
\textit{weak} multifractality.

The multifractal properties of the wave functions at criticality also
shows up in the anomalous spreading of a wave packet, where the value
of $D_2$ plays an essential role \cite{Bodo}.

The importance of the generalized dimensions and especially of $D_2$ 
therefore is obviously hard to overestimate. This is the reason why it
came as a surprise when in \cite{P-S} the authors demonstrated 
that the correlation dimension, $D_2$, has a relatively broad 
distribution even with increasing systems 
size of a three dimensional Anderson model. The authors concluded 
that this broadness remains in the thermodynamic limit and therefore 
there is no unique value of $D_2$ in this limit. Such a statement 
would imply serious objections against one--parameter scaling theory 
which is corroborated
by many other techniques \cite{rev}. Based on numerical simulations 
on a random matrix model having eigenstates that exhibit multifractality, 
however, it has been shown in \cite{E-M,EC} that the fluctuations of
$I_2$ can indeed be characterized by a universal probability distribution 
function (PDF), $\Pi(I_2)$ once $I_2$ is normalized with the typical value,
$\tilde{I}_2=I_2/I_2^{typ}$, where $I_2^{typ}=\exp\langle\ln I_2\rangle$:
\beq
\tilde\Pi(\ln \tilde{I}_2)=\Pi_N(\ln I_2/I_2^{typ} ).
\label{pi2}
\eeq
So the PDFs are simply horizontally shifted by
\beq
\ln I_2^{typ}=\langle\ln I_2\rangle\propto\tilde{D}_2\ln N
\label{i2typ}
\eeq
to put on top of each other and obtain the universal curve of
$\tilde{\Pi}(\ln\tilde{I}_2)$ showing that there is indeed a unique 
value of $\tilde{D}_2$ to characterize the scaling of $I_2$. In the 
present work we wish to reconcile the apparent
contradiction existing about the fluctuations of the correlation dimension 
$D_2$ and to show how to correct the misinterpretation
of \cite{P-S} and obtain results that are compatible with \cite{E-M,EC}.

To this end we performed numerical simulations on the power law band
random matrix ensemble that has been recently intensively investigated
\cite{E-M,EC,EC2,prbm}. This model is more suitable for a quantitative
analysis since it allows to obtain a wider range of linear extension ($N=L$) 
of the system as compared to the usual Anderson model defined over an 
$N=L^3$ lattice of linear size $L$. Moreover, the model is capable to 
show many of the
features of the traditional Anderson transition \cite{rev,VK,prbm}. We
will show that indeed the PDF of the correlation dimension of the
states is a broad function which eventually has a fixed point at a
value very close to $\tilde{D}_2$ obtained from the scaling of
$I_2^{typ}$. Moreover, from the average value, $\langle D_2\rangle$,
one can extract the same answer as well.

The elements in these ensembles of $N\times N$ complex hermitian matrices 
are Gaussian distributed random variables with zero mean, 
$\langle H_{i,j}\rangle=0$, and a variance
\beq
\langle(H_{i,j})^2\rangle=\left[1+(|i-j|/b)^2\right]^{-1}
\times\left\{\begin{array}{ll}
                    1/2     \ ,\quad & i\neq j\\
                    1/\beta \ ,\quad & i=j
       \end{array}\right.
\eeq
where $\beta=1$ or $2$ depending on the global symmetry of the system.
In the presence of time reversal symmetry, $\beta=1$, and the matrices
are real symmetric, in its absence, $\beta=2$. 
The parameter $b$ is an effective bandwidth that serves as a continuous
control parameter over a whole line of criticality \cite{VK}. In this
work we set $b=1$ as this proves to give the best convergence towards
a scale invariant distribution function \cite{E-M}. However, at the
end we will give the $b$-dependence of the results as well.

\begin{figure}
\includegraphics[width=3.4in]{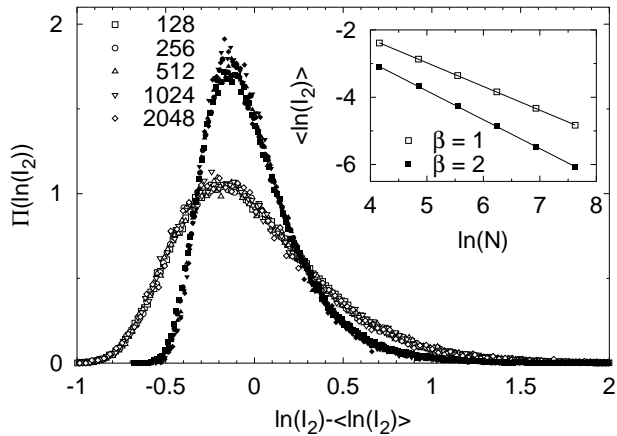}
\caption{
\label{fig1}
Distribution function of the inverse participation number $\Pi(\ln I_2)$ 
vs the rescaled value of $\ln(I_2/I_2^{typ})$ for different matrix sizes 
$N$. Open (filled) symbols correspond to $\beta=1$ ($\beta=2$).
The inset shows the scaling of the typical value, 
$\ln I_2^{typ}=\langle\ln I_2\rangle$
vs $\ln(N)$. 
}
\end{figure}

The states have been obtained using standard diagonalization routines
for the cases of $N=128,\dots,2048$. The properties of the states in
the vicinity of the middle of the band have been collected. We chose
25\% of the states around $E=0$, and verified that by narrowing this
energy window the results do not alter but obviously the statistics
becomes worse.

Each individual eigenstate may show scaling with different exponents 
$D_2$. On the other hand the distribution of the IPN, $I_2\sim N^{D_2}$,
of the eigenstates shows universal scaling properties \cite{rev}, 
thus a relation between these properties is expected. The distribution 
function of $D_2$ can be connected to that of the $I_2$ (\ref{pi2}) as
\beq
{\cal P}_N(D_2)\,dD_2=\Pi_N(I_2)\,dI_2
\eeq
which yields 
\beq
{\cal P}_N(D_2)=\Pi_N(N^{D_2})N^{D_2}\ln(N).
\label{pd2}
\eeq
We expect the distribution of $D_2$ to tend to a universal distribution 
as well as that of $I_2$. Evers and Mirlin \cite{E-M} have proved 
that in particular $I_2$ is a self--averaging quantity and numerically
confirmed that for the case of the present model. The same has been
shown for the case of the Anderson model, too \cite{EC,MEM}. Therefore 
as $N\to\infty$ we expect that the distribution 
\beq
{\cal P}_N(D_2)\to\delta (D_2-\tilde{D}_2).
\eeq
This is in direct contradiction to the conclusions of \cite{P-S}. 

\begin{figure}
\includegraphics[width=3.4in]{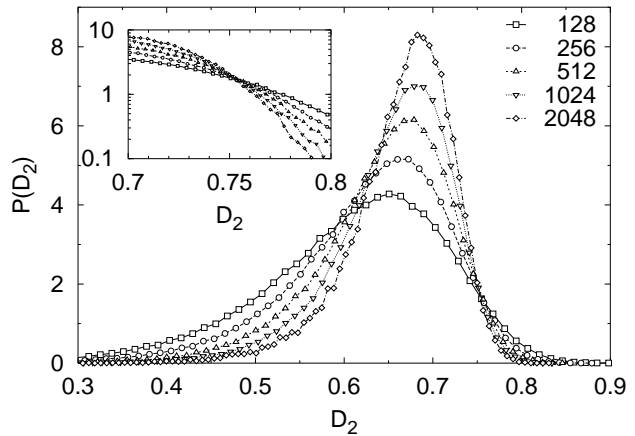}
\caption{
\label{fig2}
Distribution function of the correlation dimension ${\cal P}(D_2)$ for
the random matrix model with different matrix sizes $N$ for the case
of conserved time reversal symmetry, $\beta=1$.
}
\end{figure}

Furthermore using Eq. (\ref{pd2}) for ${\cal P}(D_2)$ and Eq. (\ref{i2typ}) 
we can readily derive expressions of the mean value and the variance of
$D_2$ in relation to the mean value and variance value of $\ln I_2$ as
\beq
\langle D_2\rangle-\tilde{D}_2=\frac{\mbox{const}}{\ln N}\to 0
\label{avd21}   
\eeq
and
\beq
\mbox{var}(D_2)=\frac{\mbox{var}(\ln I_2)}{(\ln N)^2}\to 0,
\label{vad21}
\eeq
where the convergence to zero for the variance is ensured by 
\cite{EC,EC2}
\beq
\mbox{var}(\ln I_2)=\sigma_{\infty}-aN^{-y},
\label{irrevi2}
\eeq
with the irrelevant scaling $y=\tilde{D}_2/(2\beta)>0$ and also $a>0$
and $\sigma_{\infty}>0$. These logarithmic convergences, Eqs.
(\ref{avd21}) and (\ref{vad21}), seem to be too 
difficult to detect for the case of the traditional Anderson model 
\cite{P-S}, even though, Ref. \cite{EC, EC2} showed the scaling
(\ref{irrevi2}) very clearly.

\begin{figure}
\includegraphics[width=3.4in]{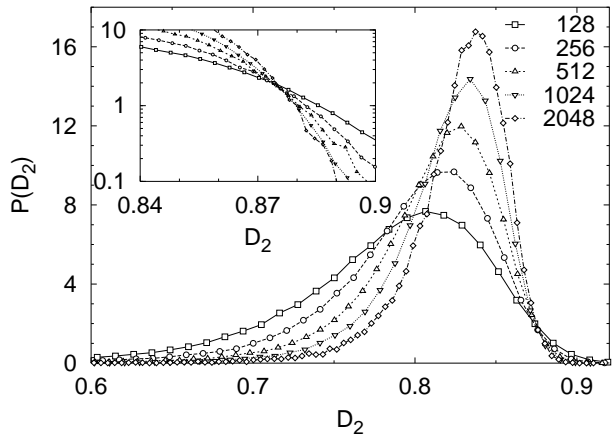}
\caption{
\label{fig3}
Distribution function of the correlation dimension ${\cal P}(D_2)$ for
the random matrix model with different matrix sizes $N$ for the case
of broken time reversal symmetry, $\beta=2$.
}
\end{figure}

In Fig.~\ref{fig1} first we show that indeed the scaling of the PDF 
$\Pi_N(\ln I_2)$ holds and obtain the value of $\tilde{D}_2$ from the
typical value, $\langle\ln I_2\rangle$ for both $\beta=1$ and $2$.
Their values are listed in Table \ref{tab}.
As a next step we turn to the PDF of the correlation dimension, $D_2$.
The correlation dimension of the individual eigenstates have been 
obtained using the standard box-counting algorithm \cite{MJ}. In 
Figs.~\ref{fig2} and \ref{fig3} we show how ${\cal P}(D_2)$ evolves by 
doubling $N$. These figures give the same information as Fig.~3 of
\cite{P-S} for the case of the power law band random matrix ensemble. By 
increasing $N$ the distribution gets narrower, moreover, there is a fixed 
point right in the interval of $0.7<D_2<0.8$ for $\beta=1$ and at 
$0.85<D_2<0.9$ for $\beta=2$. In the insets of both figures we plot 
this region enlarged in a semi-log plot. The fixed point clearly appears 
at $D_2^*\approx 0.75$ for $\beta=1$ and $D_2^*\approx 0.87$ for 
$\beta=2$. The first value is precisely the one obtained via the 
scaling of the typical value $\langle\ln I_2\rangle$ in \cite{E-M} 
and is close to our numerical results in Fig.~\ref{fig1}.

We argue that due to the self--averaging property of $I_2$ \cite{E-M}
${\cal P}(D_2)$ should tend to a $\delta$--function so for finite
$N,N'$ there should be a fixed point in the distribution function, 
\beq
{\cal P}_N(D_2^*)={\cal P}_{N'}(D_2^*), 
\eeq
provided $D_2^*$ falls within the set of $D_2$'s of the eigenstates. 
This ensures that $D_2^*=\tilde{D}_2$ is the unique value of the 
correlation dimension at $N\to\infty$. Moreover it is also the 
average value, $\langle D_2\rangle$ in this limit (\ref{avd21}).
\begin{figure}
\includegraphics[width=3.4in]{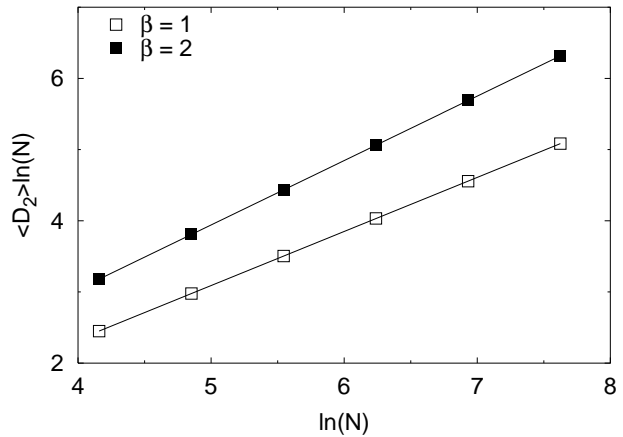}
\caption{
\label{fig4}
Scaling of the mean value of the correlation dimension as $\langle
D_2\rangle\ln(N)$ vs $\ln(N)$. The lines are linear fits to the data.
According to (\ref{avd21}) the slop corresponds to $\tilde{D}_2$
describing the scaling of $\ln I_2^{typ}$.
}
\end{figure}
The convergence of $\langle D_2\rangle$ towards $\tilde{D}_2$ is given
in Fig.~\ref{fig4}. The value of $\langle D_2\rangle$ extracted from
this fit together with the other estimates of $D_2$ are listed in
Table~\ref{tab}. 
\begin{table}
\caption[params]{\label{tab} Correlation dimensions obtained using
different approaches}
\begin{tabular}{c|cc}
                      & $\beta=1$          & $\beta=2$   \\
\hline
 $\tilde{D}_2$        & $0.7036 \pm 0.0024$ & $0.8576 \pm 0.0018$  \\
 $\langle D_2\rangle$ & $0.7598 \pm 0.0005$ & $0.9046 \pm 0.0011$  \\
 $D_2^*$              & $0.7519 \pm 0.0004$ & $0.8781 \pm 0.0014$  \\
\end{tabular}
\end{table}
The error bars characterize the goodness of the linear fits as depicted 
in Figs. \ref{fig1} and \ref{fig4}.
The different approaches give slightly different values, some of them 
coincide within a few percent difference. We attribute the discrepancies 
to possible finite size effects and to the fact that the number of samples 
has not been large enough. For conserved time reversal symmetry,
$\beta=1$, we observe a coincidence between our $\langle D_2\rangle$
and $D_2^*$ values, together with the one obtained by \cite{E-M} over
a larger ensemble of samples and a wider range of matrix size. All the 
values from different estimates seem to be within a range of 1--8\% 
difference. Therefore we may conclude that despite the relatively large 
state-to-state fluctuations of the correlation dimension depicted in a 
wide PDF of ${\cal P}(D_2)$ in Figs. \ref{fig2} and \ref{fig3} one can 
still define a unique value of $D_2$ that is characteristic to the 
critical point.

Let us apply a similar argument to the results presented in Ref.
\cite{P-S}. Inspecting Fig.~3 of that work closely, contrary to what is 
claimed there, the distribution does not show a universal, broad form but 
rather gets more and more peaked as $N$ increases. Even the authors admit 
a very large drop of ${\cal P}(D_2)$ in the interval of $1.5<D_2<2.0$ 
that apparently is system size dependent. One may find a similar fixed 
point as in our Figures \ref{fig1} and \ref{fig2} at around 
$D_2^*\approx 1.55$. This value is consistent with most of the presently 
known estimates 
of the correlation dimension calculated using different methods and
listed in \cite{P-S}. It is, however, somewhat larger then more recent 
and accurate estimates \cite{EC,MEM}. The possible reason for this
discrepancy has been discussed in detail in \cite{MEM}.

\begin{figure}
\includegraphics[width=3.4in]{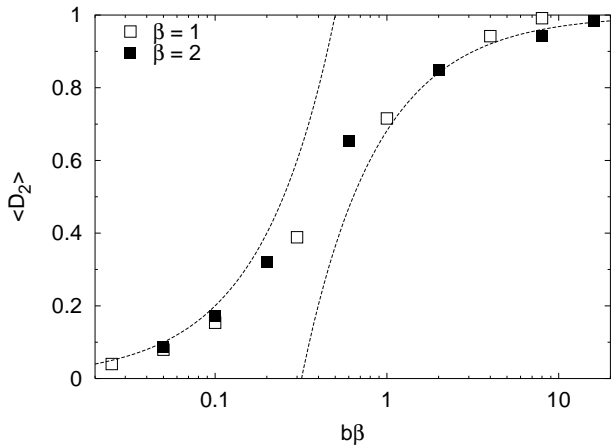}
\caption{
\label{fig5}
The $b\beta$ dependence of the correlation dimension $D_2$ obtained
from its average value. The dashed lines correspond to the analytical
estimates in the limiting cases of $b\ll 1$ and $b\gg 1$ (\ref{d2an}).
}
\end{figure}

Finally we present the behavior of the correlation dimension
as a function of the parameter $b$. This is presented in Fig.~(\ref{fig5})
which shows that in fact $D_2$ is a unique function of $b\beta$. In
this figure we plotted the slope of $\langle D_2\rangle\ln N$ vs $\ln
N$ for the various values of $b$ and $\beta$. The
analytical estimates of \cite{E-M}
\beq
D_2=\left\{\begin{array}{ll}
                    1-(\pi b\beta)^{-1} \ ,\qquad & b\gg 1\\
                    2b\beta \ ,\qquad & b\ll 1.
           \end{array}\right.
\label{d2an}
\eeq
According to the figure we see that our numerical simulations are in
full accordance with the analytical estimates. The discrepancies here
also are attributed to the considerably smaller set of samples and
smaller range of matrix size $N$ as compared to \cite{E-M}.

In conclusion we have performed numerical simulations of the power law
band random matrices in order to obtain the PDF of the correlation
dimension. We found that this function has a fixed point, $D_2^*$, very
close to the value of $\tilde{D}_2$ obtained from the scaling of the
typical IPN, $I_2^{typ}$. Besides a direct relation between the
average correlation dimension, $\langle D_2\rangle$, enabled us to
obtain a third estimate of the correlation dimension. All of these are
in good agreement with each other that corroborates the expectation,
that there is a unique value of the correlation dimension
characterizing the states at the Anderson transition. For finite size 
simulations, the state-to-state fluctuations of both the IPN and the 
correlation dimension are large, nevertheless, both the PDF of $I_2$
and that of $D_2$ provide a unique value of $\tilde{D}_2$ in the limit
$N\to\infty$. This way we have reconciled the apparent contradiction
between the state-to-state fluctuation of the correlation dimension
and the universal scaling of the IPN.
Similar results may be achieved for other generalized dimensions, e.g. 
the information dimension, $D_1$ \cite{vi2}.

The author thanks financial support from the Alexander von Humboldt 
Foundation and from the OTKA Nos. T032116 and T034832.

{}

\end{document}